\documentclass{ws-mod}

\newcommand{\aem    }{\mbox{$\alpha$}}
\newcommand{\aemsq  }{\mbox{$\aem^2$}}
\newcommand{\epem   }{\mbox{$\rm e^+e^-$}}
\newcommand{\qsq    }{\mbox{$Q^{2}$}}
\newcommand{\psq    }{\mbox{$P^{2}$}}
\newcommand{\lamq   }{\mbox{$\Lambda_{\rm QCD}^{2}$}}
\newcommand{\ft     }{\mbox{$F_{2}^{\gamma}$}}
\newcommand{\fl     }{\mbox{$F_{\rm L}^{\gamma}$}}
\newcommand{\ftxq   }{\mbox{$\ft(x,\qsq)$}}
\newcommand{\flxq   }{\mbox{$\fl(x,\qsq)$}}
\newcommand{\ftp    }{\mbox{$F_{2}^{p}$}}
\newcommand{\fteff  }{\ensuremath{F_{eff}^{\gamma}}}
\newcommand{\avqsq  }{\mbox{$\langle\qsq\rangle$}}
\newcommand{\gevsq  }{\mbox{$\rm GeV^2$}}

\begin{document}

\title{Measurements of hadronic structure functions of the photon at LEP}

\author{R. J. Taylor}

\address{Department of Physics and Astronomy, University College London,\\
Gower Street, London WC1E 6BT, United Kingdom\\
E-mail: rjt@hep.ucl.ac.uk}


\maketitle

\abstracts{
The present status of the measurements of hadronic structure functions
of the photon, investigated in deep inelastic electron-photon scattering
at LEP, is presented.
This article covers the hadronic structure function \ft\ of quasi-real
photons as well as the structure function \fteff\ of virtual photons.
Special emphasis is given to new developments in the analysis and to the
most recent measurements.}

\section{Introduction}

The photon is unique in that it can act both as a fundamental particle - the
gauge boson of QED - or as an extended object with structure.
The structure function of the photon differs from that of the proton because
the photon has a point-like coupling to quark charges - which is calculable
in perturbative QCD - as well as a non-perturbative hadron-like part.

The classic way to investigate the structure of the photon is via the deep
inelastic scattering of electrons (or positrons) on the quasi-real photons 
which accompany the beams at \epem\ colliders. 
The structure function \ft, which in leading order is
proportional to the sum over the parton densities of the photon weighted by
the square of the parton's charge, 
can be extracted by measuring the differential cross section
for this process and through recourse to the following relation:
 \begin{equation}
  \frac{\mbox{d}^2\sigma_{\rm e\gamma\rightarrow e X}}{\mbox{d}x\mbox{d}Q^2}
   =\frac{2\pi\aemsq}{x\,Q^{4}}
  \left[\left( 1+(1-y)^2\right) \ftxq - y^{2} \flxq\right]
 \label{eqn:Xsect}
 \end{equation}
where $\qsq=-q^2$ is the negative value of the four-momentum squared 
of the virtual probe photon, \aem\ is the fine structure 
constant and $x$ and $y$ are the usual dimensionless 
deep inelastic scattering variables.
In the kinematic region explored at LEP, $y^2\ll 1$ and the term proportional 
to \fl\ in Equation~\ref{eqn:Xsect} can be neglected.

This article concentrates on the most recent results on photon structure from
the LEP collaborations. A recent, comprehensive review can be found in 
Reference 1, which also contains the references to all LEP results in addition
to those explicitly referenced here.

\section{Recent improvements in the analysis}

Because the electron which emits the quasi-real photon is not seen in the
detector, $x$ needs to be determined by measuring the invariant mass, $W$,
of the
hadronic final state. The hadrons tend to be boosted in the forward direction
and thus poorly contained in the detector,
so a central component of structure function analyses at LEP is the use
of an unfolding procedure to relate the visible distributions to the underlying
ones.

The unfolding requires the input of a reference Monte Carlo model and this
leads to a dependence of the \ft\ measurement on the modelling of the hadronic
final state.
As a consequence, past LEP measurements of \ft\ have suffered from
large model dependent errors, and significant effort has been invested in
exploring techniques to reduce these.

In order to stimulate improvements in the available Monte Carlo models,
ALEPH, L3 and OPAL have combined their data and produced corrected 
distributions of variables related to the hadronic final state~\cite{ALO}.
The differences between experiments are used to estimate the systematic errors,
and these errors are usually smaller than the significant differences between 
models. This result thus serves as a useful constraint on the development
of the models.

ALEPH and OPAL have employed the method of two-dimensional unfolding~\cite{2D}
to reduce the sensitivity of the result to the large differences observed in 
the modelling of the hadronic final state. This relies on the fact that 
discrepancies between data and Monte Carlo in the variable in which you unfold
are not important. There is information in every event about the angular
distribution of the hadrons, but this information is not exploited if only $x$
is used in the unfolding. If a variable that characterises the final state
hadrons is chosen and used as a second unfolding variable, the result will be
independent of the differences between Monte Carlo models in that variable as
well as in $x$, thus reducing the overall model dependence.

Improving the reconstruction of $W$ will also reduce the uncertainties due to
the models, and several techniques have been employed to do this.
Both OPAL and L3 have used a method which uses information from the tagged 
electron together with transverse momentum conservation~\cite{J-B}. This
improves the reconstruction because the resolution on electromagnetic
energy measurements is generally better than those on hadronic energy. 
OPAL have also introduced a special treatment of the energy in the forward
region to make the detector response more uniform~\cite{PR314}.
For their analysis of \ft\ at high \qsq, L3 have made use of a kinematic fit,
and showed this to give a good correlation between the generated and
measured $W$~\cite{L3}. 

In the past, QED radiative corrections have been neglected in measurements of
\ft. However, the most recent OPAL analyses~\cite{PR314}$^,$\cite{mine} have 
corrected for the effects of initial state radiation and the Compton scattering
process using the RADEG program~\cite{RADEG}. The radiative corrections are 
$x$-dependent and largest at small values of $x$, such that the shape of \ft\ 
is changed when the corrections are applied.

\begin{figure}[b]
%\figurebox{20pc}{15pc}{} % to have a box alone
\begin{center}
\epsfxsize=21pc % will enlarge or reduce the postscript figures based on the xsize
\epsfbox{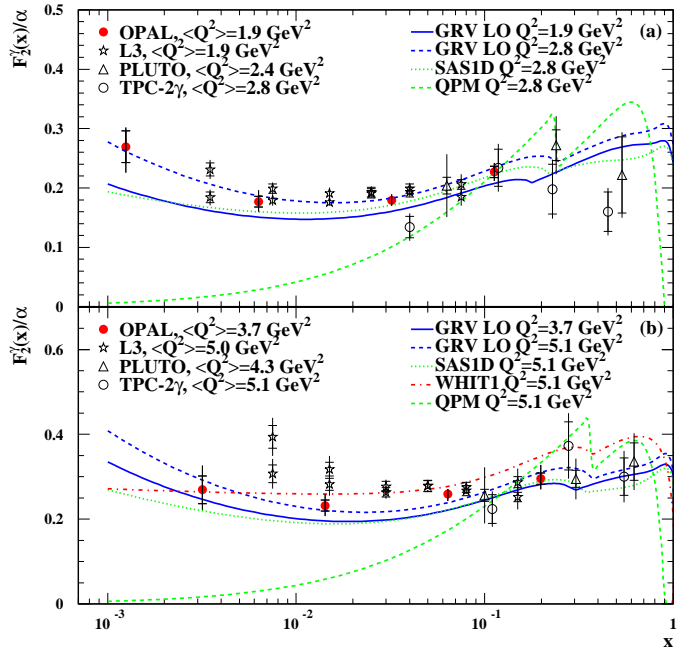} % postscript image file name
\end{center}
\caption{The measurements of \ft\ at low $x$   \label{fig:PR314_20}}
\end{figure}

\section{Recent results}

The rise in \ftp\ at low $x$ seen at HERA has precipitated intensive 
investigation to try and establish whether such a rise is also present in \ft,
and LEP is able to extend the reach at low $x$ relative to
measurements at previous \epem\ colliders. 
OPAL has recently published measurements~\cite{PR314} 
concentrating on the behaviour of \ft\ at low $x$ for \avqsq\ between 1.9 
and 17.8 \gevsq, examples of which are shown in Figure~\ref{fig:PR314_20} along
with a number of earlier measurements. The systematic uncertainties have been
considerably reduced compared to previous analyses, through the use of the 
techniques described in the previous section, along with improved Monte Carlo 
models. In addition, account has been taken of the fact that when
systematic errors are evaluated by modifying parameters in the analysis a
significant statistical component can be introduced if, for example, events
are added to or removed from bins. A procedure has been used to evaluate the
size of this component, which is then subtracted from the systematic error.

The shapes of the GRV LO and SaS1D parameterisations are generally consistent
with the OPAL and L3 data. The data are consistent with, but do not
conclusively prove, the presence of a rise in \ft\ at low $x$. They do,
however, demonstrate that the photon must contain a significant hadron-like
component at low $x$.

\begin{figure}[htb]
%\figurebox{20pc}{15pc}{} % to have a box alone
\begin{center}
\epsfxsize=15pc % will enlarge or reduce the postscript figures based on the xsize
\epsfbox{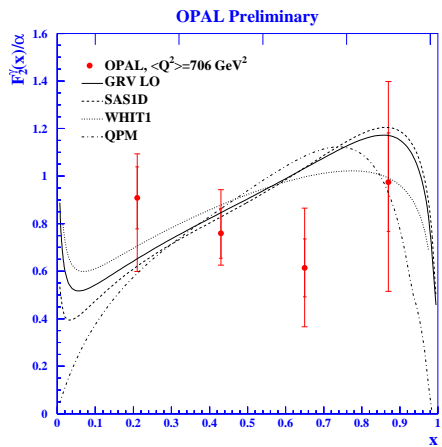} % postscript image file name
\end{center}
\caption{New OPAL measurement of \ft\ at high \qsq.  \label{fig:pn454_05}}
\end{figure}

At high \qsq\ the point-like part of \ft\ is expected to dominate.
A new OPAL measurement of \ft\ at the highest \avqsq\ thus far~\cite{mine} 
is shown in Figure~\ref{fig:pn454_05}. At large \qsq\ the hadronic system
has more transverse momentum and as a consequence there is much better 
correlation between the true and measured invariant mass. This leads to a much
smaller dependence of the result on the unfolding and the input Monte Carlo
model. The present measurement suffers from large statistical and systematic
uncertainties, which should be reduced once the final analysis - which will
include the full LEP2 data sample - is published.

\section{Virtual photon structure}

A suppression of \ft\ is expected with increasing virtuality of the target
photon, \psq, although the various theoretical predictions show a large spread
in the magnitude of this suppression. For real photons only transverse
helicity states of the target photon contribute. For $\psq \gg 0$, though, 
longitudinal helicities must in addition be taken into account, and 
Equation~\ref{eqn:Xsect} is not valid. However, if the condition 
$\qsq \gg \psq \gg \lamq$ is satisfied, an effective structure function 
\fteff\ can be defined.

\begin{figure}[thb]\unitlength 1pt
\begin{center}
{\includegraphics[width=0.49\linewidth,clip]{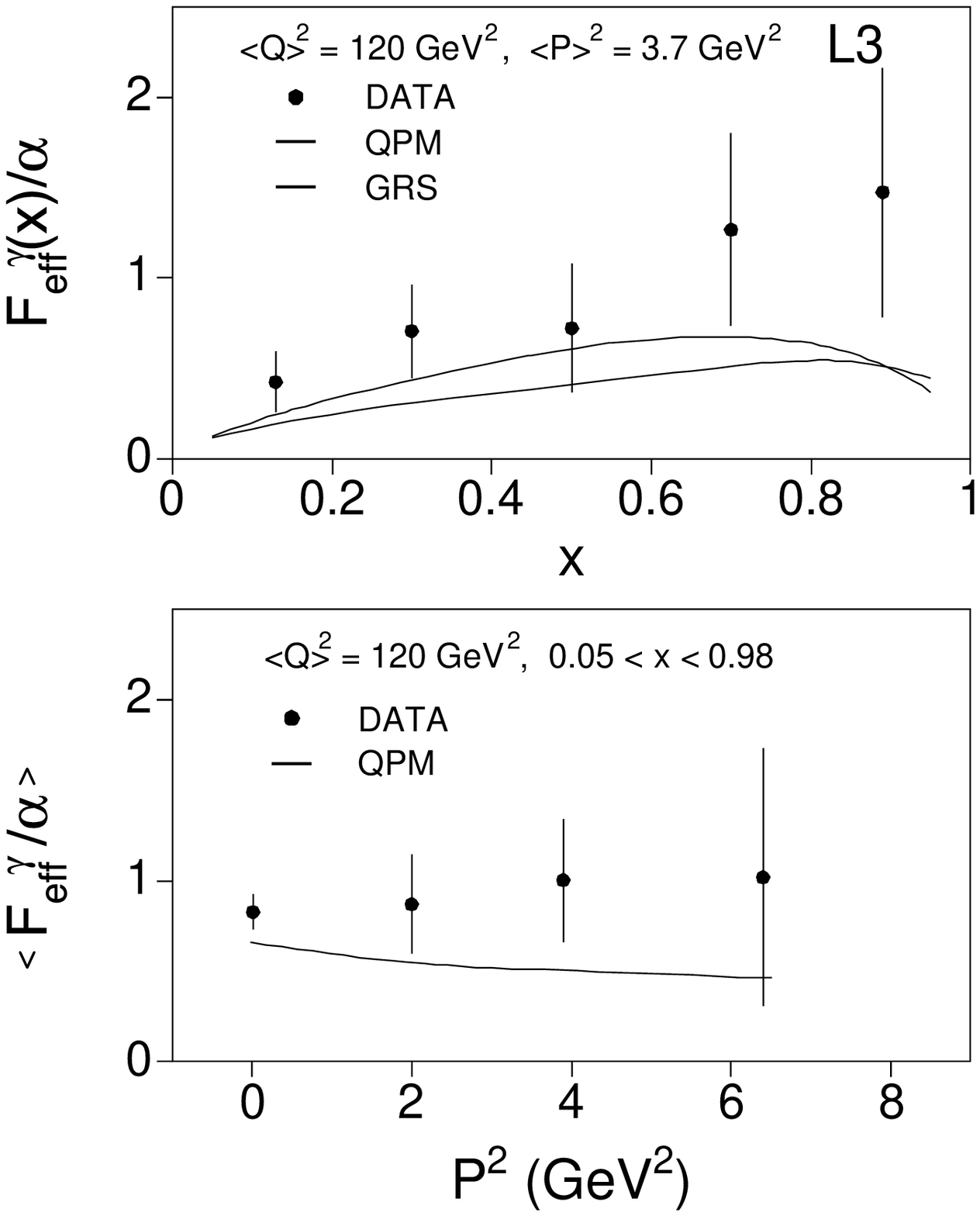}}
{\includegraphics[width=0.49\linewidth,clip]{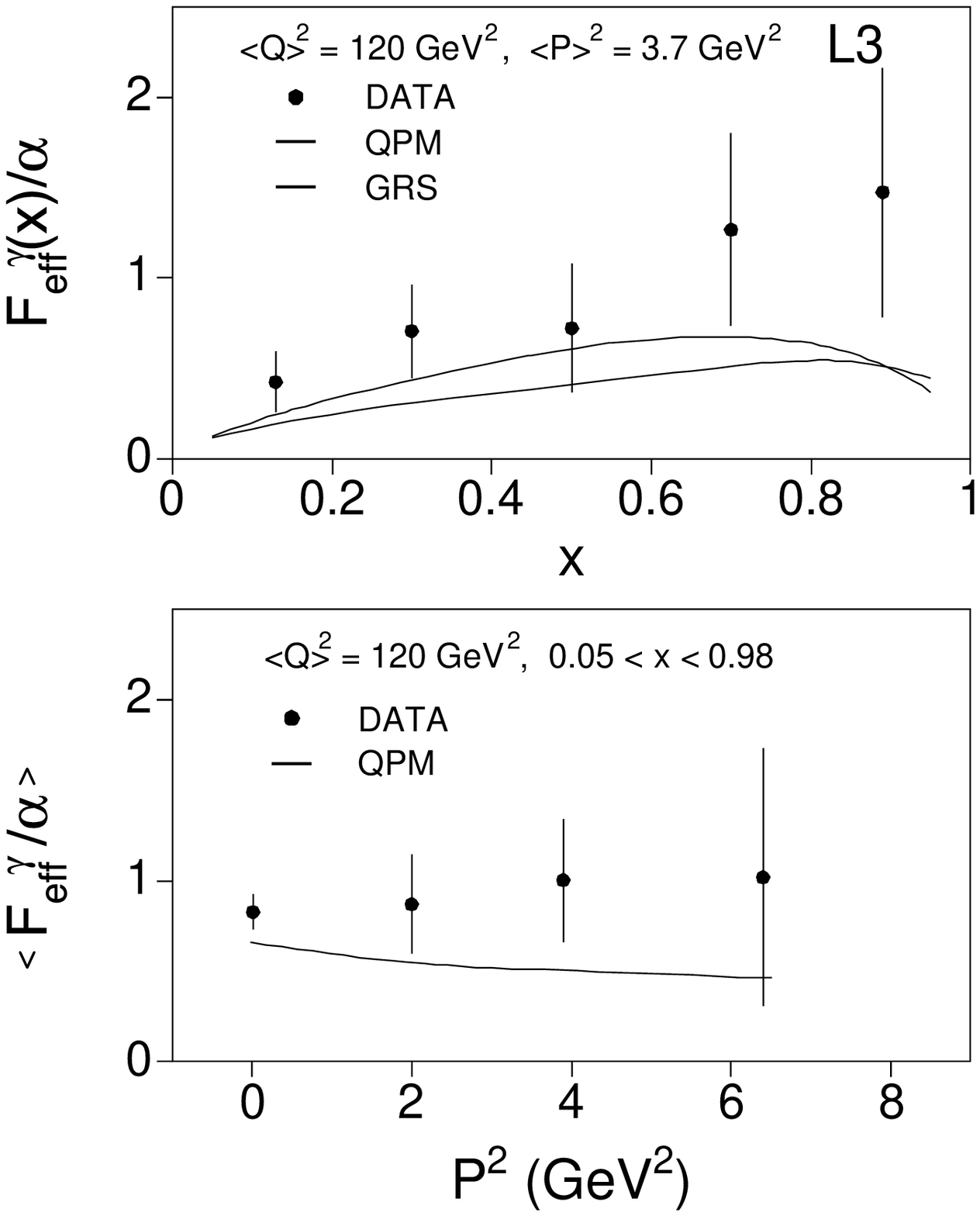}}
\caption{\label{fig:L3}
L3 measurement of \fteff.
        }
\end{center}
\end{figure}

Only L3 have measured \fteff\ at LEP~\cite{L3} and their result is shown in 
Figure~\ref{fig:L3}. Both theoretical predictions undershoot the data -
something that is to be expected since the QPM does not include the hadron-like
contribution and the GRS parameterisation takes account only of transverse
virtual photons. The measured \psq\ dependence is consistent in shape with 
the QPM prediction, but the statistics are severely limited and the full LEP2
data need to be exploited to shed further light on this aspect of the photon's
structure.
%-----------------------------------------------------------------------
%-----------------------------------------------------------------------
\vspace*{-6mm}
\section{Summary and Conclusions}

\begin{figure}[htb]
%\figurebox{20pc}{15pc}{} % to have a box alone
\begin{center}
\epsfxsize=21pc % will enlarge or reduce the postscript figures based on the xsize
\epsfbox{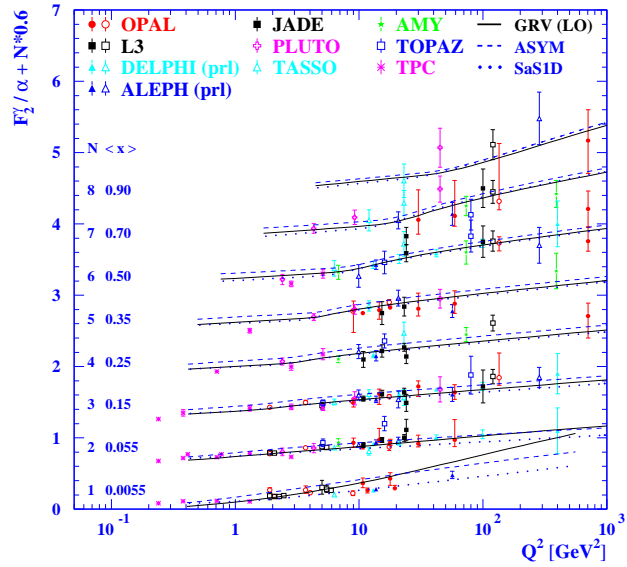} % postscript image file name
\end{center}
\caption{Summary of the measurements of the \qsq\ evolution of \ft in bins of $x$ (taken from Reference 1). \label{fig:summ}}
\end{figure}

LEP has considerably improved our understanding of the hadronic structure
of the photon. Recently, it has been demonstrated that the systematic
uncertainties can be reduced through methods such as two
dimensional unfolding. The current status is shown in Figure~\ref{fig:summ},
which shows the evolution of \ft\ with \qsq. One can see the positive scaling
violations for all values of $x$ which result from the point-like coupling.

Despite recent advances, however, further improvements, in particular in the
available Monte Carlo models, are desirable. Also, many results are still
statistically limited and the exploitation of the full LEP2 data will 
enhance our understanding of the hadronic structure of the
photon.

\end{document}